\documentclass{PoS}
\usepackage{amsmath}
\usepackage{epstopdf}
\newcommand{\Nf}{N_{\text{f}}}

\newcommand{\Ns}{N_{\text{s}}}
\newcommand{\Nt}{N_{\text{t}}}

\newcommand{\sz}{\sigma_0}
\newcommand{\cx}{C(x)}
\newcommand{\ctk}{\tilde{C}(k)}
\newcommand{\ii}{\ensuremath{\mathrm{i}}}
\title{Lattice investigation of the phase diagram of the 1+1 dimensional
	Gross-Neveu model at finite number of fermion flavors
}
\ShortTitle{Phase diagram of the 1+1 dimensional Gross-Neveu model at finite $\Nf$}
\author{\speaker{Laurin Pannullo}$^a$, Julian Lenz$^{b}$, Marc Wagner$^a$,
		Bj\"orn Wellegehausen$^{b}$, Andreas Wipf$^{b}$\\
        \llap{$^a$}Goethe Universit\"at Frankfurt am Main, Institut f\"ur Theoretische Physik,\\Max-von-Laue-Stra{\ss}e 1, D-60438 Frankfurt am Main, Germany\\
        \llap{$^b$}Theoretisch-Physikalisches-Institut, Friedrich-Schiller-Universit\"at Jena,\\ Fr\"obelstieg 1, D-07743 Jena, Germany\\     
        E-mail: \email{pannullo@itp.uni-frankfurt.de}, \email{julian.johannes.lenz@uni-jena.de}, \email{mwagner@itp.uni-frankfurt.de}, \email{bjoern.wellegehausen@uni-jena.de}, \email{wipf@tpi.uni-jena.de}}
\abstract{We explore the phase structure of the 1+1 dimensional Gross-Neveu model at finite number of fermion flavors using lattice field theory. Besides a chirally symmetric phase and a homogeneously broken phase we find evidence for the existence of an inhomogeneous phase, where the condensate is a spatially oscillating function. Our numerical results include a crude $\mu$-$T$ phase diagram.}
\FullConference{37th International Symposium on Lattice Field Theory - Lattice2019\\
		16-22 June 2019\\
		Wuhan, China}
\begin{document}


\section{Introduction, the Gross-Neveu model}

Exploring the QCD phase diagram using lattice field theory is currently limited to rather small chemical potential due to the sign problem (see e.g. \cite{Philipsen:2010gj,Aarts:2015tyj}). Thus, it is of interest to study the phase structure of simpler models, which have similarities to QCD at least in certain aspects. A common example is the Gross-Neveu (GN) model \cite{Gross:1974jv}.

The Euclidean action and the partition function of the GN model in 1+1 spacetime dimensions are
\begin{align}
S = \int \mathrm{d}^2x \, \bigg(\sum_{n=1}^{\Nf} \bar{\psi}_n \bigg(\gamma_0 (\partial_0 + \mu) + \gamma_1 \partial_1\bigg) \psi_n - \frac{g^2}{2} \bigg(\sum_{n=1}^{\Nf} \bar{\psi}_n \psi_n\bigg)^2\bigg) \quad , \quad Z = \int \mathrm{D}\bar{\psi} \, \mathrm{D}\psi \, e^{-S} ,
\end{align}
where $\psi_n$ denotes a fermionic field with flavor index $n = 1,\ldots,\Nf$, $\mu$ is the chemical potential and $g$ is the coupling constant. A possible irreducible representation for the $\gamma$ matrices, which we use throughout this work, is $\gamma_0 = \sigma_1$ and $\gamma_1 = \sigma_2$. To get rid of the four-fermion interaction, one typically introduces a real scalar field $\sigma$. Integrating over the fermionic fields then leads to the equivalent effective action and partition function
\begin{align}
S_\textrm{eff} = \Nf \bigg(\frac{1}{2 \lambda} \int \mathrm{d}^2x \, \sigma^2 - \ln\Big(\det\Big((\partial_0 + \mu) \gamma_0 + \partial_1 \gamma_1 + \sigma\Big)\Big)\bigg) \quad , \quad Z = \int \mathrm{D}\sigma \, e^{-S_\textrm{eff}} ,
\end{align}
where $\lambda = \Nf g^2$.

The effective action $S_\text{eff}$ has a discrete symmetry, $S_\text{eff}[\sigma]=S_\text{eff}[-\sigma]$. One can show $\langle \sigma \rangle \propto \langle \sum_{n=1}^{\Nf} \bar{\psi}_n \psi_n \rangle$, where $\langle \ldots \rangle$ denotes the path integral expectation value. Thus, a non-vanishing $\langle \sigma \rangle$ would indicate spontaneous breaking of the symmetry $\psi_n \rightarrow \sigma_3 \psi_n$. Since $\sigma_3$ anticommutes with $\gamma_0$ and $\gamma_1$, it is appropriate to define $\gamma_5 = \sigma_3$ and to interpret the symmetry $\psi_n \rightarrow \sigma_3 \psi_n = \gamma_5 \psi_n$ as discrete chiral symmetry and $\langle \sigma \rangle$ as corresponding order parameter.


\section{\label{SEC001}The Gross-Neveu model for $\Nf \rightarrow \infty$}

\begin{figure}[htb]
	\centering
	\includegraphics[width=12cm]{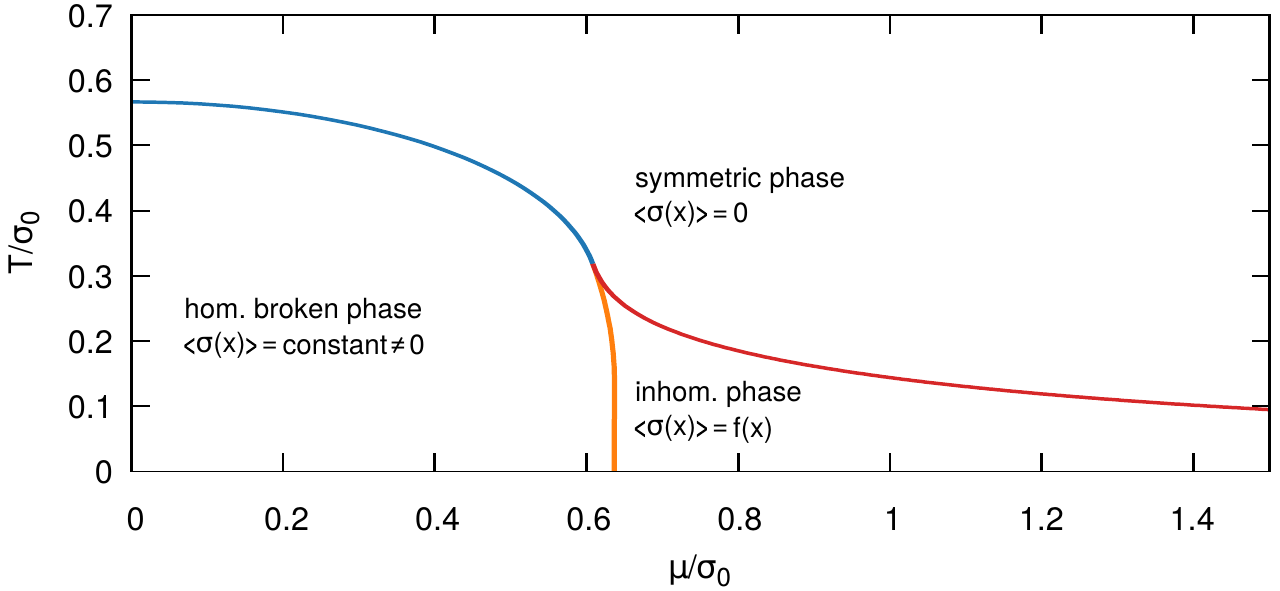}
	\caption{\label{Fig:largeNphase}Phase diagram of the GN model in in 1+1 spacetime dimensions for $\Nf \rightarrow \infty$ \cite{Thies:2003kk}.}
\end{figure}

In the limit $\Nf \rightarrow \infty$ the phase diagram of the GN model in 1+1 spacetime dimensions was calculated analytically \cite{Thies:2003kk,Schnetz:2004vr}, and with lattice field theory and related numerical methods \cite{deForcrand:2006zz,Wagner:2007he,Heinz:2015lua}. There are three phases as shown in Figure~\ref{Fig:largeNphase}:
\begin{itemize}
	\item \emph{a homogeneously broken phase} ($\langle \sigma(x) \rangle = \text{constant} \neq 0$),

	\item \emph{a symmetric phase} ($\langle \sigma(x) \rangle = 0$),

	\item \emph{an inhomogeneous phase} ($\langle \sigma(x) \rangle = f(x)$, where $f(x)$ is an oscillating periodic function of the spatial coordinate).
\end{itemize}
In the inhomogeneous phase close to the phase boundary to the homogeneously broken phase $f(x)$ has a kink-antikink shape. For increasing chemical potential the wavelength of $f(x)$ and its amplitude decrease and the shape is similar to a sin-function. For a recent review on inhomogeneous condensates see ref.\ \cite{Buballa:2014tba}.

\section{The Gross-Neveu model at finite $\Nf$}

In this work we performed lattice field simulation with chiral fermions to explore the phase structure of the GN model in 1+1 spacetime dimensions at finite $\Nf$ with particular focus on the possible existence of an inhomogeneous phase (see also ref.\ \cite{Pannullo:2019bfn} for results obtained at an early stage of this project). We simulated a large variety of ensembles for various values of $\Nf$, $\mu$, the number of lattice sites $\Nt \times \Ns$ and the lattice spacing $a$ (the latter is a function of the coupling constant $\lambda$ and can be adjusted, by tuning $\lambda$ appropriately). The temperature is proportional to the inverse temporal extent of the lattice, $T = 1 / \Nt a$, and the spatial extent of the lattice is denoted by $L = \Ns a$. We employed two different fermion discretizations, naive fermions and SLAC fermions (see e.g.\ \cite{Wozar:2011gu}), which provides a powerful cross-check of our numerical results.

We set the scale via the absolute value of the chiral condensate at $\mu = 0$ and $T = 0$, \\ $\sz = \langle | \sigma | \rangle\rvert_{\mu=0, T=0}$. Dimensionful quantities are either expressed in units of the lattice spacing $a = 1$ (e.g.\ $x \equiv x/a$, $\mu \equiv \mu a$, $T \equiv T a$) or in units of $\sz$ (e.g.\ $x \sz$, $\mu/\sz$, $T/\sz$).


\subsection{The aligned condensate $\Sigma(x)$}

The expectation value $\langle \sigma(x) \rangle$ is not a suitable quantity, to reliably detect a possibly existing inhomogeneous phase. This is so, because field configurations might exhibit similar periodic oscillations, but could be spatially shifted relative to each other. Thus, more appropriate is the ``aligned condensate'' $\Sigma(x)$, where field configurations are matched by spatial translations $x \rightarrow x - \Delta x$, before the ensemble average is computed,
\begin{align}
\Sigma(x) = \frac{1}{\Nt} \sum_{t=0}^{\Nt} \Big\langle \sigma(t,x-\Delta x) \Big\rangle .
\end{align}
The technical aspects of the matching of field configurations, i.e.\ the determination of $\Delta x$ for each field configuration, will be discussed in detail in an upcoming publication.

\begin{figure}[htb!]
	\centering
	\includegraphics[width=14cm]{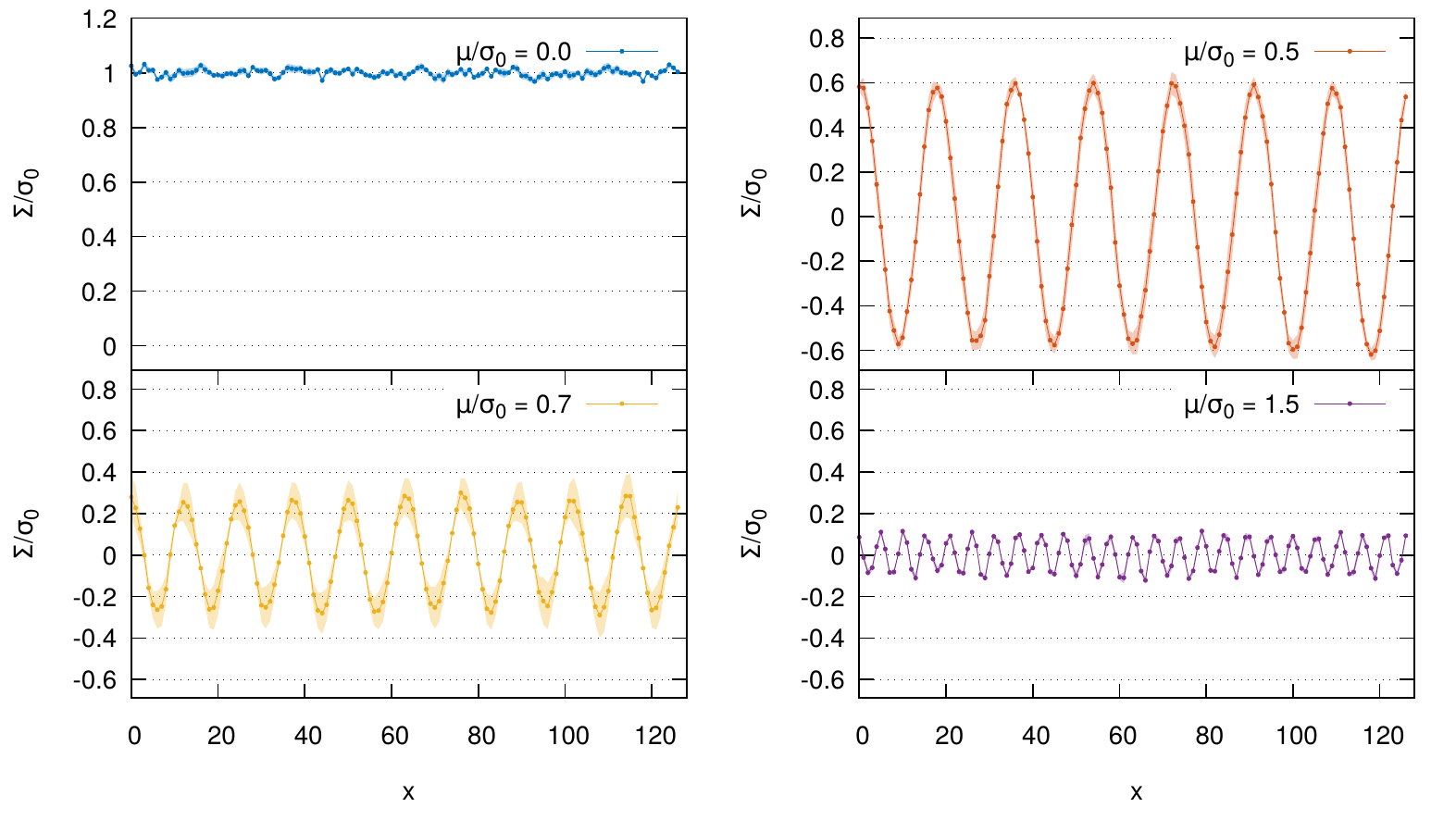}
	\caption{\label{Fig:Sigma}Aligned condensate $\Sigma(x)$ for $\Nf = 8$, $\mu/\sz \in \{ 0.0 , 0.4 , 0.6 , 0.8 \}$ and low temperature $T/\sz = 0.031$ (SLAC fermions, $\lambda = 5.20$, $\Ns = 127$, $\sz = 0.408$).}
\end{figure}

Exemplary results for $\Sigma(x)$ for $\Nf = 8$, $\mu/\sz \in \{ 0.0 , 0.4 , 0.6 , 0.8 \}$ and low temperature \\ $T/\sz = 0.031$ are presented in Figure~\ref{Fig:Sigma}. These results are quite similar to $\langle \sigma(x) \rangle$ for $\Nf \rightarrow \infty$ as discussed in section~\ref{SEC001}. The aligned condensate $\Sigma(x)$ is a non-vanishing constant for $\mu = 0$. For larger chemical potential $\Sigma(x)$ starts to oscillate, which signals the existence of an inhomogeneous phase. When increasing $\mu$ even further, both the wavelength and the amplitude of $\Sigma(x)$ become smaller.

\subsection{The correlation function $\cx$}

Further quantities to distinguish a possibly existing inhomogeneous phase from a homogeneously broken phase or a symmetric phase is the correlation function
\begin{align}
\cx = \frac{1}{\Nt \Ns} \sum_{t,y} \Big\langle \sigma(t,y+x) \sigma(t,y) \Big\rangle
\end{align}
and its Fourier transform
\begin{align}
\ctk = \frac{1}{\Ns} \sum_x \exp\bigg(-\frac{2 \pi \ii k x}{\Ns}\bigg) \cx \quad , \quad k \in \left\{ - \frac{\Ns}{2}, \ldots , + \frac{\Ns}{2} -1 \right\} .
\end{align}
Their expected behavior is as follows:
\begin{itemize}
	\item Inside a homogeneously broken phase:
	\\ $\cx$ will approach a constant $\approx \sz^2$ exponentially fast. $\ctk$ will exhibit a pronounced peak around $k = 0$.

	\item Inside a symmetric phase:
	\\ $\cx$ will approach zero exponentially fast. $\ctk$ for $k \neq 0$ is similar as in a homogeneously broken phase, however, without a peak around $k = 0$.

	\item Inside an inhomogeneous phase:
	\\ $\cx$ will oscillate with constant, non-vanishing amplitude for large separations $x$. $\ctk$ will exhibit pronounced peaks around $\pm k_\text{peak}$. The wavelength of the oscillations of $\cx$ is $\approx \Ns/k_\text{peak}$.
\end{itemize}

\begin{figure}
	\centering
	\includegraphics[width=14cm]{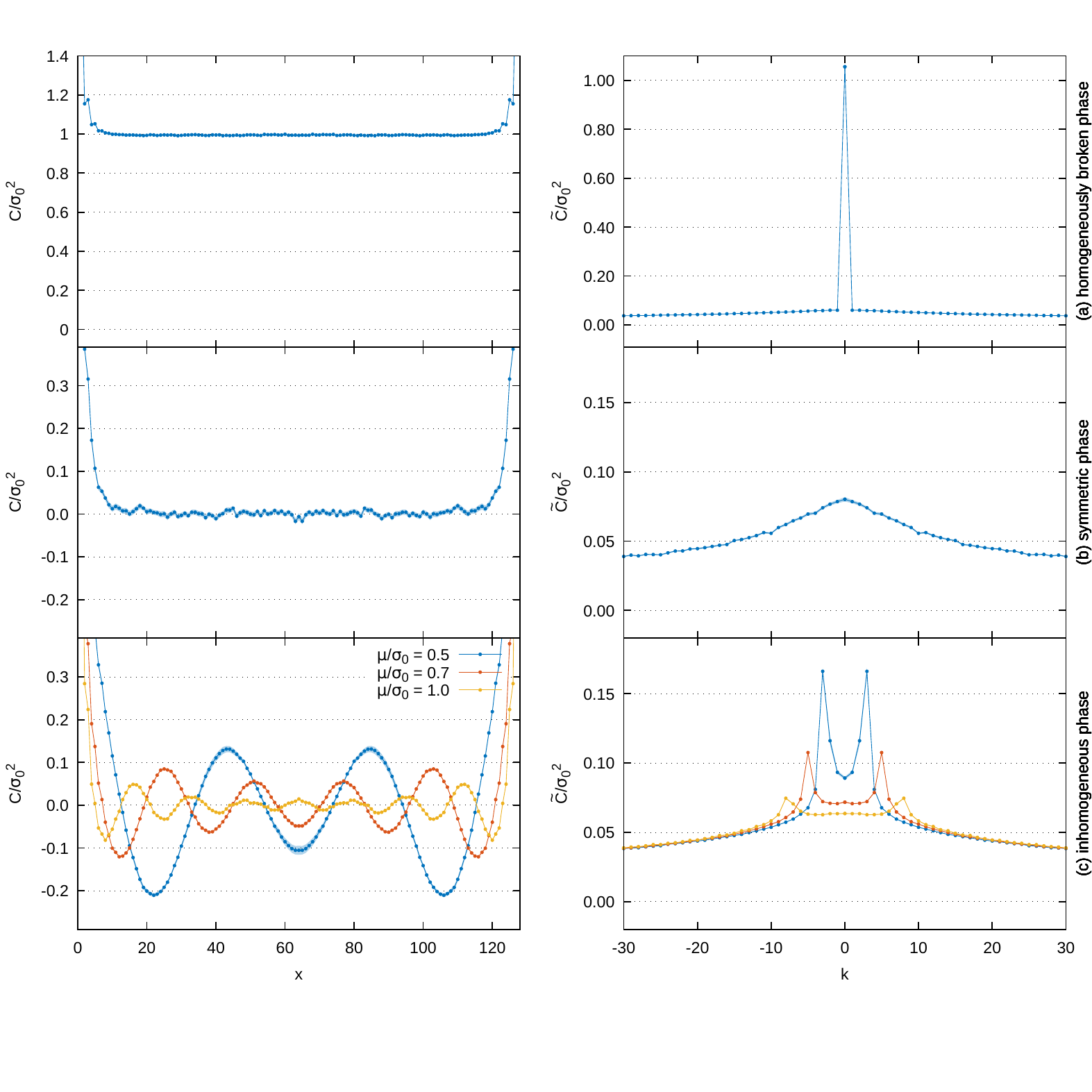}
	\caption{\label{Fig:Cx}Correlation functions $\cx$ and $\ctk$ for $\Nf = 8$ (Naive fermions, $\lambda = 5.28$, $\Ns = 128$, $\sz = 0.2015$). \textbf{(a)}~Inside a homogeneously broken phase, $\mu/\sz = 0.0$, $T/\sz = 0.062$. \textbf{(b)}~Inside a symmetric phase, $\mu/\sz = 0.0$, $T/\sz = 0.621$. \textbf{(c)}~Inside an inhomogeneous phase, $\mu/\sz \in \{0.5 , 0.7 , 1.0\}$, $T/\sz=0.078$. $\ctk$ is not plotted in the whole range of k. }
\end{figure}

Exemplary results for $\cx$ and $\ctk$ for $\Nf = 8$ are presented in Figure~\ref{Fig:Cx} and agree with this expectation. Particularly interesting are the two plots at the bottom showing $\cx$ and $\ctk$ at large chemical potential $\mu/\sz \in \{0.5 , 0.7 , 1.0\}$ and low temperature $T/\sz=0.078$ inside an inhomogeneous phase. For increasing $\mu$, both the wavelength and the amplitude become smaller, as in the limit $\Nf \rightarrow \infty$ (see section~\ref{SEC001}).

\begin{figure}
	\centering
	\includegraphics[width=14cm]{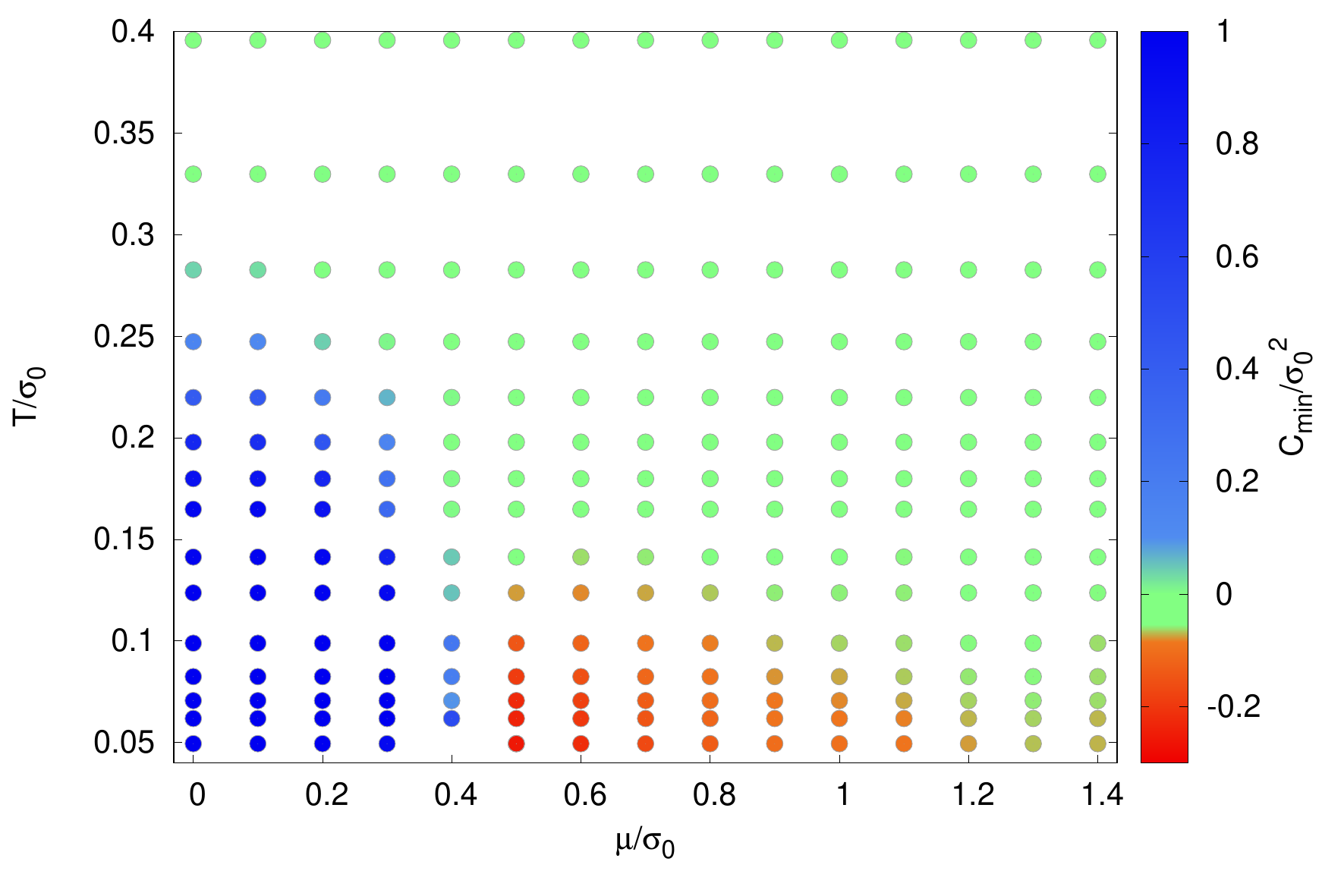}
	\caption{\label{Fig:Cmin}Phase diagram via $\min_x \, \cx$ for $\Nf = 8$ (Naive fermions, $\lambda = 4.84$, $\Ns = 128$, $\sz = 0.253$).}
\end{figure}

From Figure~\ref{Fig:Cx} one can see
\begin{align}
	\min_x \, \cx \begin{cases}
	> 0       & \text{inside a homogeneously broken phase}\\
	\approx 0 & \text{inside a symmetric phase}\\
	< 0       & \text{inside an inhomogeneous phase}\\
	\end{cases}.
\end{align}
Thus, the minimum of the correlation function $\cx$ is suited to plot a crude phase diagram, as e.g.\ shown in Figure~\ref{Fig:Cmin} for $\Nf = 8$. This phase diagram is qualitatively similar to the $\Nf \rightarrow \infty$ phase diagram in Figure~\ref{Fig:largeNphase} (blue = homogeneously broken phase, green = symmetric phase,
red = inhomogeneous phase). The homogeneously broken phase and the inhomogeneous phase are, however, somewhat smaller for finite $\Nf$ than for $\Nf \rightarrow \infty$.

All results shown in this section were obtained from lattices with
either $\Ns = 127$ or $\Ns = 128$ sites in spatial direction. To
investigate and exclude finite volume corrections, we are currently in
the process of performing identical simulations with lattices of even
larger spatial extent, $\Ns = 512$, at $\Nf=2$. We obtain similar results, which we will discuss in detail in an upcoming publication.

\acknowledgments

We thank M.\ P.\ Lombardo for very helpful discussions and suggestions.
L.P.\ thanks A.\ Sciarra for helpful discussions and support in preparing this conference talk.
L.P.\ thanks the organizers of the ''The 37th International Symposium on Lattice Field Theory'' for the possibility to give this talk.
L.P.\ and M.W.\ acknowledge support by the Deutsche Forschungsgemeinschaft (DFG, German Research Foundation) through the CRC-TR 211 ``Strong-interaction matter under extreme conditions'' -- project number 315477589 -- 
TRR 211.
M.W.\ acknowledges support by the Heisenberg Programme of the Deutsche Forschungsgemeinschaft, grant WA 3000/3-1.
This work was supported in part by the Helmholtz International Center for FAIR within the framework of the LOEWE program launched by the State of Hesse.
Calculations on the GOETHE-HLR and on the FUCHS-CSC high-performance computer of the Frankfurt University were conducted for this research. We would like to thank HPC-Hessen, funded by the State Ministry of Higher Education, Research and the Arts, for programming advice.
Calculations on the local HPC cluster of the FS-University Jena were conducted for this research.



\end{document}